\documentclass{aa}
\usepackage{graphicx}
\usepackage{txfonts}
\usepackage{xcolor}
\newcommand{\flag}[1]{\texttt{#1}}

\begin{document}

   \title{K2 results for "young" $\alpha$-rich stars in the Galaxy}


   \author{V. Grisoni
          \inst{1,2}
          \and
          C. Chiappini\inst{3}
                    \and
          A. Miglio\inst{1,4, 5}
                    \and        
          K. Brogaard\inst{1,6}
                    \and
          G. Casali\inst{1,4}
                    \and
          E. Willett\inst{5}
                   \and
         J. Montalb\'an\inst{1} 
                    \and \\
          A. Stokholm\inst{1,4,5,6} 
          \and
        J. S. Thomsen\inst{1,4,6}
         \and
        M. Tailo\inst{1} 
        \and
        M. Matteuzzi\inst{1,4} 
         \and
         M. Valentini\inst{3}          
         \and
         Y. Elsworth\inst{5,6}
         \and
         B. Mosser\inst{7}
          }

  \institute{
  {Department of Physics \& Astronomy, University of Bologna, Via Gobetti 93/2, 40129 Bologna, Italy}\label{difa}
  \and
{INAF, Osservatorio Astronomico di Trieste, via G.B. Tiepolo 11, I-34131, Trieste, Italy}\\\email{valeria.grisoni@inaf.it}
\and
{Leibniz-Institut fur Astrophysik Potsdam (AIP), An der Sternwarte 16, D-14482 Potsdam, Germany} \label{aip}
\and
{INAF – Osservatorio di Astrofisica e Scienza dello Spazio, Via P. Gobetti 93/3, 40129 Bologna, Italy} \label{oas}
\and
{School of Physics and Astronomy, University of Birmingham, Edgbaston, Birmingham, B15 2TT, UK}
\label{bhm}
  \and
  {Stellar Astrophysics Centre, Department of Physics \& Astronomy, Aarhus University, Ny Munkegade 120, 8000 Aarhus C, Denmark}\label{aarhus}
    \and
{LESIA, Observatoire de Paris, Universit\'e PSL, CNRS, Sorbonne Universit\'e, Universit\'e de Paris, 92195 Meudon, France} \label{lesia}
}

   \date{Received --; accepted --}

 
  \abstract
   {
The origin of apparently young $\alpha$-rich stars in the Galaxy is still a matter of debate in Galactic archaeology, whether they are genuinely young or might be products of binary evolution and merger/mass accretion.}
   {We aim to shed light on the nature of young $\alpha$-rich stars in the Milky Way by studying their distribution in the Galaxy thanks to an unprecedented sample of giant stars that cover different Galactic regions and have precise asteroseismic ages, chemical, and kinematic measurements.}
   {We analyze a new sample of $\sim$ 6000 stars with precise ages coming from asteroseismology. Our sample combines the global asteroseismic parameters
measured from light curves obtained by the K2 mission with stellar parameters and chemical abundances obtained from APOGEE DR17 and GALAH DR3, then
cross-matched with \textit{Gaia} DR3. We define our sample of young $\alpha$-rich stars and study their chemical, kinematic, and age properties.}
   {We investigate young $\alpha$-rich stars in different parts of the Galaxy and we find that the fraction of young $\alpha$-rich stars remains constant with respect to the number of high-$\alpha$ stars at $\sim10\%$. Furthermore, young $\alpha$-rich stars have kinematic and chemical properties similar to high-$\alpha$ stars, except for [C/N] ratios. }
{Thanks to our new K2 sample, we conclude that young $\alpha$-rich have similar occurrence rate in different parts of the Galaxy and they share similar properties as the normal high-$\alpha$ population, except for [C/N] ratios. This suggests that these stars are not genuinely young, but products of binary evolution and merger/mass accretion. Under that assumption, we find the fraction of these stars in the field to be similar to that found recently in clusters. This fact suggests that $\sim$10$\%$ of the low-$\alpha$ field stars could also have their ages underestimated by asteroseismology. This should be kept in mind when using asteroseismic ages to interpret results in Galactic archaeology.}
   {}

   \keywords{Galaxy: abundances --
                Galaxy: formation --
                Galaxy: evolution -- stars: late-type –asteroseismology
               }

   \maketitle
%

\section{Introduction}

The goal of Galactic archaeology is to unveil the history of formation and evolution of the Galaxy from abundance patterns, kinematics and stellar ages \citep[for a recent review, see][]{Matteucci2021}.
\\We are in an era of great advances for this field of research thanks to the advent of large spectroscopic surveys, such as Apache Point Observatory Galactic Evolution Experiment \citep[APOGEE,][]{Majewski2017}, \textit{Gaia}-ESO Survey \citep[GES,][]{Gilmore2012}, Large Sky Area Multi-Object Fiber Spectroscopic Telescope \citep[LAMOST,][]{Cui2012}, Radial Velocity Experiment \citep[RAVE,][]{Steinmetz2006}, Radial Velocity Spectrometer \citep[RVS,][]{RVS}, GALactic Archaeology with HERMES \citep[GALAH,][]{DeSilva2015}, that can provide detailed stellar abundances and radial velocities of stars in the Milky Way. This information combined with \textit{Gaia} mission \citep[]{Gaia2016} can be used to obtain the full 6D phase space information for large samples of stars \citep[e.g.][among others]{Queiroz2023}. Furthermore, these surveys can then be combined with missions such as COnvection ROtation and planetary Transits \citep[CoRoT,][]{Baglin2006}, {\it Kepler} \citep{Gilliland2010}, Transiting Exoplanet Survey Satellite \citep[TESS,][]{Ricker2014} and K2 \citep{Howell2014} that allow us to infer precise stellar ages through asteroseismology, offering novel perspectives to the study of the formation and evolution of the Milky Way \cite[see e.g.][]{Miglio2009,Miglio2013,Casagrande2016,Anders2017a,Pins2018,SilvaAguirre2018,Valentini2019,Rendle2019,Warfield2021,Miglio2021,Montalban2021, Mackereth2021,Zin2022,Stello2022}.
\\The Milky Way disc shows two distinct sequences in the [$\alpha$/Fe] vs. [Fe/H] diagram 
(\citealt{Fuhrmann1998}, and more recently \citealt{Bensby2014,Anders2014,RecioBlanco2014,Mikolaitis2014,Mikolaitis2017,Hayden2015,RojasArriagada2017,Queiroz2020}). Stars of the high-$\alpha$ sequence are generally older than stars of the low-$\alpha$ one 
\citep{Haywood2013,Bensby2014,SilvaAguirre2018,Miglio2021}, and the presence of these two sequences can be interpreted by means of detailed models of Galactic chemical evolution, with the high-$\alpha$ sequence forming on a shorter timescale with respect to the low-$\alpha$ one 
\citep[e.g.][]{Chiappini1997,Chiappini2009,Grisoni2017,Grisoni2021,Spitoni2019,Spitoni2021}. Thus, in general, the [$\alpha$/Fe] ratio is considered to be a relevant indicator to understand galaxy formation and evolution \citep[see][]{Matteucci2001,Matteucci2012}. Although the interpretation of the discontinuity between the high-$\alpha$ and low-$\alpha$ sequences varies \citep[e.g.][]{Buck2020,Khoperskov2021}, there is a consensus that the high-$\alpha$ population is dominated by older stars \citep{Haywood2013,Bensby2014}, as also recently confirmed by asteroseismology \citep{SilvaAguirre2018,Miglio2021,Montalban2021}. 
\\However, by exploring a data set combining spectroscopy from APOGEE and asteroseismology from CoRoT (CoRoGEE), \cite{Chiappini2015} reported the discovery of a group of young [$\alpha$/Fe]-enhanced stars in their sample. These young [$\alpha$/Fe]-enhanced stars (hereafter, young $\alpha$-rich stars) appeared of particular interest since they could not be explained by classical chemical evolution models: they presented high [$\alpha$/Fe] values like the high-$\alpha$ stars, but asteroseismic young ages ($<$8 Gyr) in contrast to what we would expect from the models. One possible interpretation of the young $\alpha$-rich stars is that they might be products of mass transfer / merger: they have higher mass and therefore they appear young when explained by single-star evolutionary theory. In this scenario, they should be present in every direction where the high-$\alpha$ population extends. However, these young $\alpha$-rich stars seemed to be more abundant towards the inner Galactic disc regions and, therefore, \cite{Chiappini2015} suggested a second interpretation in which the origin of these stars could be related to Galactic evolution, and in particular to the peculiar chemical evolution that occurs near the corotation region of the Galactic bar. In a companion paper, \cite{Martig2015} added the discovery of a group of young (massive) [$\alpha$/Fe]-rich stars in the \textit{Kepler} field. The presence of young $\alpha$-rich-stars has been detected also by other studies using different methods for age determination \citep[see][]{Haywood2013,Bensby2014,Bergemann2014}.
\\With a radial-velocity-monitoring campaign,  \cite{Jofre2016} concluded that a large fraction of their sample of young $\alpha$-rich stars could be binaries \citep[see also more recently][]{Jofre2022}. \cite{Izzard2018} performed a detailed theoretical study and, by means of a binary population-nucleosynthesis model, showed that it is possible to have young $\alpha$-rich stars through a binary channel. Therefore, young $\alpha$-rich stars can offer a new way to infer (close) binary fractions in the Galaxy.

\begin{figure}
    \centering
	\includegraphics[scale=0.45]{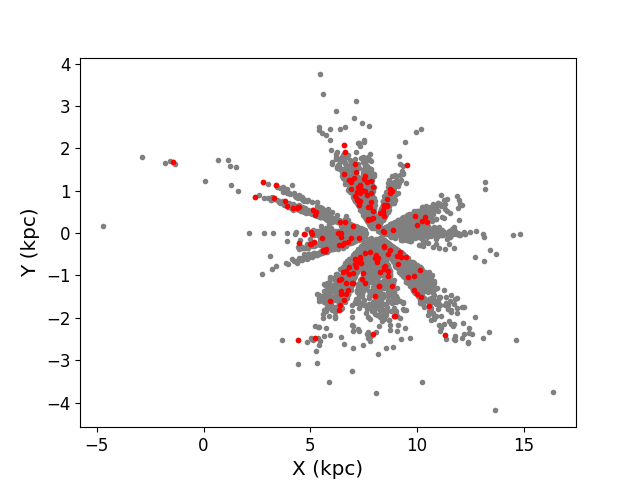}
 	\includegraphics[scale=0.45]{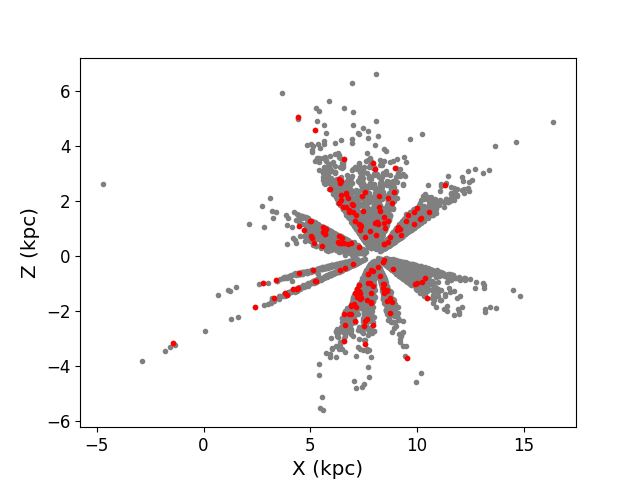}
   	\includegraphics[scale=0.45]{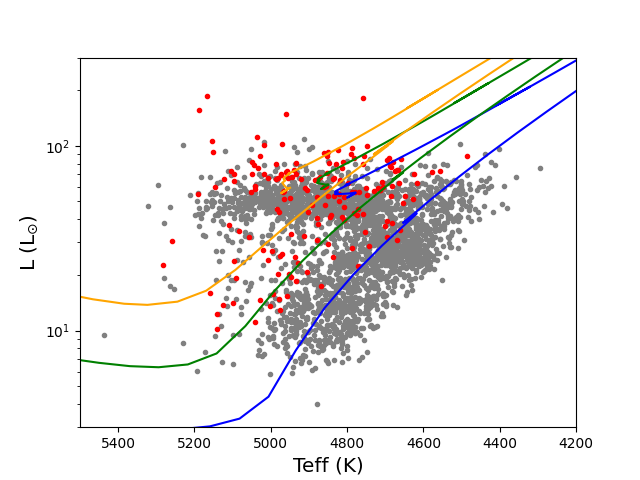}
    \caption{Upper and middle panels: Location of the K2 sample considered in this work in Galactocentric coordinates: (X,Y) and (X,Z) planes, respectively. The whole sample is in gray, and the young-young-$\alpha$ rich stars are represented with red dots. Lower panel: Hertzsprung–Russell diagram for stars in our sample, compared with stellar tracks from \cite{Miglio2021} with [m/H]=-0.25 and different masses (1 M$_{\bigodot}$ in blue, 1.4 M$_{\bigodot}$ in green, 1.8 M$_{\bigodot}$ in orange). The high-$\alpha$ stars are in gray, and the young-$\alpha$ rich stars are represented with red dots. 
    }
    \label{Fig1}
\end{figure}

From the kinematic point of view, the young $\alpha$-rich stars show kinematic properties similar to the high-$\alpha$ sequence \citep{SilvaAguirre2018,Sun2020,Zhang2021}, thus suggesting that these young $\alpha$-rich stars might be similar to the high-$\alpha$ sequence from the point of view of Galaxy evolution. Also from the chemical abundance point of view, \cite{Yong2016} and \cite{Matsuno2018} have found that the young $\alpha$-rich stars in general have properties similar to the high-$\alpha$ sequence. \cite{Hekker2019} included CNO elements in their analysis and found anomalies that could be interpreted as clues of mergers or mass transfer scenario, since those abundances can be affected by stellar evolution effects \citep[e.g.][]{Salaris2015}. Similar conclusions were also reached by \cite{Sun2020} and \cite{Zhang2021}, by performing a detailed chemical and kinematic analysis of a sample of young $\alpha$-rich stars in the LAMOST survey, where they concluded that young $\alpha$-rich stars should not be considered really young, but rather as the product of binary evolution. More recently, \cite{Cerqui2023} by using APOGEE abundances and astroNN ages supported the idea that young $\alpha$-rich stars might be considered as stragglers of the high-$\alpha$ population \citep[see also][]{Jofre2022}. In this context, asteroseismology can then provide a powerful method to provide precise stellar ages and complement the results from other methods for age determination.
\\Alternative ideas suggesting these stars to be really young have been presented in the recent literature. \cite{Weinberg2017} and \cite{Johnson2020} suggested that young $\alpha$-rich stars could arise from bursts of star formation that enhance the rate of core-collapse supernovae (SNe) enrichment. Conversely, \cite{Johnson2021} proposed that those stars are not so much "$\alpha$-rich" but rather "Fe-poor" because of less Type Ia SNe events: thus, they explained a population of young and intermediate-age $\alpha$-enhanced stars as caused by migration-induced variability in the occurance rate of SNe Ia. Recently, \cite{Borisov2022} studied lithium, masses, and kinematics of young Galactic dwarf and giant stars with extreme [$\alpha$/Fe] ratios and concluded that, at variance with \cite{Zhang2021}, those stars should be considered as effectively young: however, they left open the interpretation about their origin and suggested that their high [$\alpha$/Fe] ratio might reflect massive star ejecta in recent enhanced star formation episodes in the Galactic thin disc, for example due to interactions with the Sagittarius dwarf galaxy.
\\Recently, \cite{Miglio2021} investigated the occurrence of young $\alpha$-rich stars in their \textit{Kepler} sample and supported the scenario in which most of these overmassive stars had experienced interaction with a companion. Furthermore, an occurrence rate of overmassive red giant stars of about 10$\%$ has been also found in recent studies of open clusters \citep{Handberg2017,Brogaard2018,Brogaard2021}.
In the last years, several other studies have confirmed the presence of young $\alpha$-rich stars in their samples \citep{SilvaAguirre2018,Wu2018,Wu2019,Sun2020,Ciuca2021,Queiroz2023,Cerqui2023}.
Still, thus, the origin of young $\alpha$-rich stars is debated in the field of Galactic archaeology, whether they are really young or their apparent young ages are due to binary interactions or merger events. 
\\The aim of this paper is to investigate the origin of young $\alpha$-rich stars by taking advantage of a new sample with available asteroseismic data from K2 mission, astrometric information from \textit{Gaia} and abundances from APOGEE DR17 and GALAH DR3. Our new K2 sample allows to extend previous asteroseismic studies by covering larger spatial baselines and with a much larger sample than before \citep[see e.g. the CoRoGEE study by][]{Chiappini2015}; moreover, it allows to complement other recent works on young $\alpha$-rich stars using similar chemical abundances from APOGEE survey, but very different methods for age determination \citep[see e.g.][]{Jofre2022,Cerqui2023}. In this way, we can account for precise stellar ages from asteroseismology and take advantage from a sample that span a wider range of Galactocentric distances in order to perform a novel comprehensive analysis of young $\alpha$-rich stars in the Galaxy. The paper is organized as follows. In Section 2, we describe our sample, with its asteroseismic, spectroscopic and astrometric constraints. In Section 3, we define the young $\alpha$-rich stars in our sample and discuss their properties. Finally, in Section 4, we summarize our main conclusions.

\section{Stellar sample}


\begin{table*}
	\centering
	\caption{Young $\alpha$-rich stars in different Galactic regions. In the first column, we indicate the considered range in guiding radius $R_g$ (in kpc) and Galactic height |Z| (in kpc). In the second, third and fourth columns, we indicate the corresponding total number of stars, high-$\alpha$ (HA) stars and young $\alpha$-rich (YAR) stars, respectively. In the fifth and sixth columns, we show the fraction of young $\alpha$-rich stars with respect to the total of stars and with respect to the high-$\alpha$ sequence, respectively. We report our results for ages computed with ($\nu_{\textup{max}}$, L) and, in parentheses, with ($\nu_{\textup{max}}$, $\Delta \nu$).}
	\label{Tab1}
	\begin{tabular}{lcccccr} 
		\hline
		$R_g$ (kpc) & TOT & HA & YAR & f$_{TOT}$& f$_{HA}$\\
		\hline
		$<$5 & 419 (284) & 296 (202) & 23 (18)  & 5.6$^{+1.0}_{-1.9}$\% (6.3$^{+2.1}_{-1.4}$\%) & 7.8$^{+1.4}_{-1.4}$\% (8.9$^{+2.9}_{-1.9}$\%) \\
		5-7 & 1943 (1630) & 1208 (981) & 81 (106) & 4.2$^{+0.4}_{-0.3}$\%  (6.5$^{+0.6}_{-0.6}$\%)  & 6.7$^{+0.6}_{-0.5}$\% (10.8$^{+0.9}_{-0.9}$\%)  \\
  	    7-8 & 1326 (1194) & 444 (374) & 26 (38)  & 2.0$^{+0.5}_{-0.2}$\% (3.2$^{+0.5}_{-0.5}$\%)  & 5.9$^{+1.4}_{-0.5}$\% (10.2$^{+1.6}_{-1.6}$\%)  \\
    	8-10 & 2347 (2096) & 469 (389) &  28 (34) & 1.2$^{+0.2}_{-0.1}$\% (1.6$^{+0.4}_{-0.2}$\%)  & 6.0$^{+1.0}_{-0.6}$\% (8.7$^{+2.0}_{-1.0}$\%) \\
		$>$10  & 838 (681) & 111 (66)  &  10 (10)  & 1.2$^{+0.2}_{-0.2}$\%   (1.5$^{+0.4}_{-0.4}$\%)  &  9.0$^{+1.8}_{-1.8}$\%  (15.2$^{+4.5}_{-4.5}$\%) \\
		\hline
	\end{tabular}
 \\
 	\begin{tabular}{lcccccr} 
		\hline
		|Z| (kpc)  & TOT & HA & YAR & f$_{TOT}$& f$_{HA}$\\
		\hline
		$<$0.5 & 1759 (1646) & 195 (178)  & 18 (23) &  1.0$^{+0.1}_{-0.1}$\% (1.4$^{+0.2}_{-0.3}$\%) & 9.2$^{+0.5}_{-0.5}$\% (12.9$^{+2.2}_{-2.8}$\%)  \\
		0.5-1 & 2573 (2318)  & 670 (595) & 35 (58)  & 1.4$^{+0.2}_{-0.2}$\% (2.5$^{+0.3}_{-0.2}$\%)  & 5.2$^{+0.9}_{-0.7}$\%  (9.7$^{+1.2}_{-1.0}$\%)  \\
		$>$1  & 2541 (1921) & 1663 (1239)  & 115 (125)  &   4.5$^{+0.5}_{-0.2}$\%  (6.5$^{+0.6}_{-0.5}$\%)  & 6.9$^{+0.7}_{-0.3}$\%  (10.0$^{+1.0}_{-0.8}$\%) \\
		\hline
	\end{tabular}
\end{table*}

\begin{figure*}
    \centering
	\includegraphics[scale=0.35]{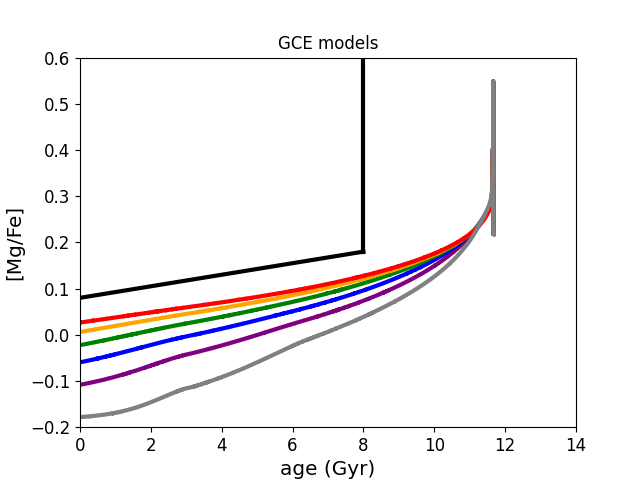}
	\includegraphics[scale=0.35]{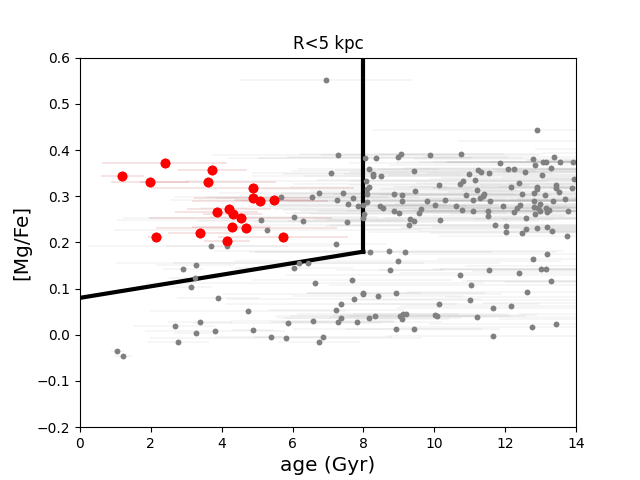}
 	\includegraphics[scale=0.35]{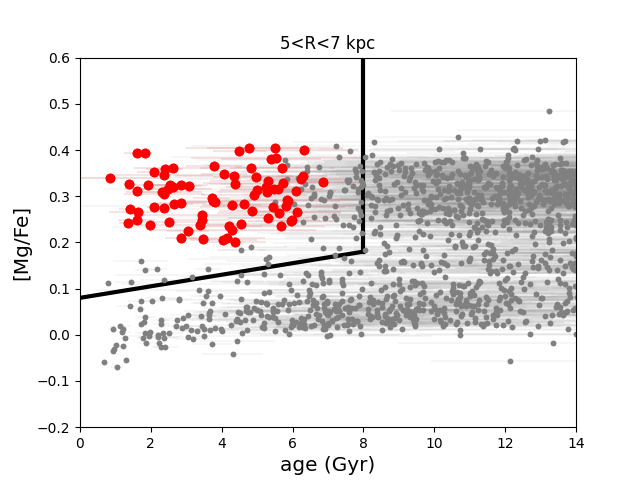}
    \includegraphics[scale=0.35]{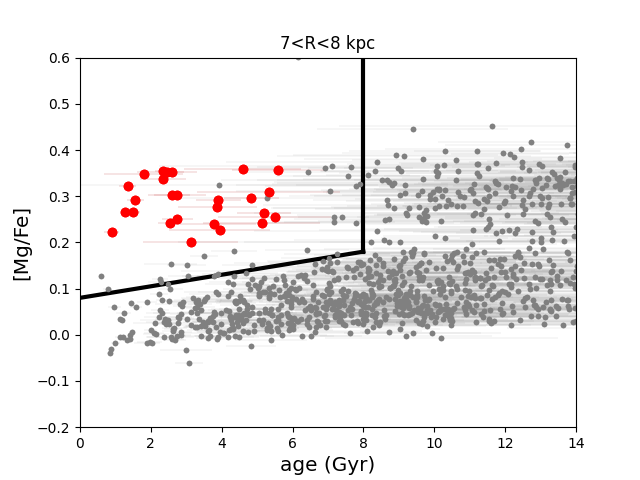}
   	\includegraphics[scale=0.35]{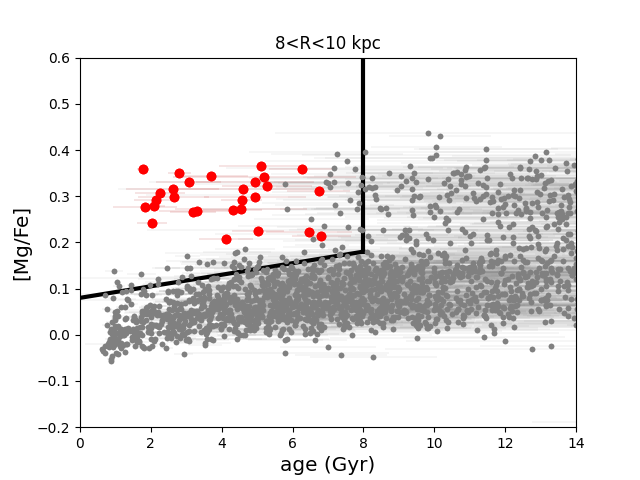}
    \includegraphics[scale=0.35]{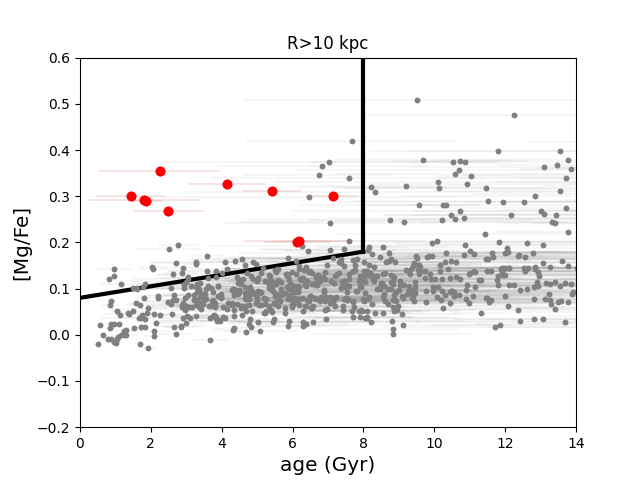}
    \caption{[Mg/Fe] vs. age diagrams. In the upper left, we show the predictions of models of the thin disc at different radii: 4 kpc in gray, 6 kpc in purple, 8 kpc in blue, 10 kpc in green, 12 kpc in orange, 14 kpc in red  \citep[see][]{Chiappini2009,Grisoni2017,Grisoni2018}. In the other panels, we show the observational data for our sample in different bins of guiding radius. The black lines in the various plots mark the 'forbidden region' as defined in \cite{Chiappini2015}. Red larger dots are the young $\alpha$-rich stars selected in our sample. We report the 1 $\sigma_{age}$ error bars. 
    }
    \label{Fig3}
\end{figure*}

Our observational sample combines the global asteroseismic parameters measured from light curves obtained with K2 mission \citep{Howell2014} with chemical abundances inferred from high-resolution spectra taken by the APOGEE DR17 survey \citep{APOGEEDR17} and also GALAH DR3 \citep{Buder2021}, and astrometric information from the Third Data Release of \textit{Gaia} (\textit{Gaia} DR3, \citealt{GaiaDR3}).
\subsection{Asteroseismic, spectroscopic and astrometric constraints}
The sample considered in this work (see Willett et al. in prep.) is obtained across the campaigns 1 – 8 and 10 – 18 of the K2 mission \citep{Howell2014}. Although the K2 mission provides only 80d light curves at variance with $\sim$4yr light curves from \textit{Kepler}, K2 data have the advantage of offering a wide coverage across different parts of the Galaxy, as can be seen from Fig.~\ref{Fig1}. Here, we take into account global asteroseismic parameters, in particular the frequency of maximum oscillation power ($\nu_{\textup{max}}$) and the average large frequency separation ($\Delta\nu$) from the pipeline of \cite{Elsworth2020}. In this context, we remove targets with low $\nu_{\textup{max}}$ ($<20$ $\mu$Hz), where asteroseismic scaling relations have not been well tested so far. 
\\
Our targets were observed also by the APOGEE and GALAH surveys. The chemical abundances used here are those given by the APOGEE Stellar Parameters and Chemical Abundances Pipeline (ASPCAP; \citealt{ASCAP2016}). A full description of this pipeline as applied to APOGEE DR17 is given in Holtzman et al. (in prep). We discard targets with \flag{ASPCAPFLAG} having \flag{STAR\_BAD} or \flag{STAR\_WARN}, and with \flag{RV\_FLAG}.
We complement our analysis by using GALAH data for neutron-capture elements not available in APOGEE. Regarding GALAH, we consider the chemical abundances from GALAH DR3 \citep{Buder2021}. In this case, we consider only the elemental abundances with \flag{flag\_X\_fe}$==$0 \citep[see][]{Buder2021} for the neutron-capture elements considered in this work.
\\
We then use astrometric information from \textit{Gaia} DR3 \citep{GaiaDR3}. We discard targets with \flag{ruwe}$>$1.4 or which are marked as binaries by \flag{non\_single\_stars} flag \citep{Lindegren2018}. 
\subsection{Inferring stellar ages and orbital parameters}
Masses, radii, ages and distances are inferred by using the code PARAM \citep{DaSilva2006,Rodrigues2014,Rodrigues2017}, which makes use of a Bayesian inference method. PARAM takes as input a combination of asteroseismic indices and spectroscopic constraints, such as $\nu_{\textup{max}}$, [Fe/H], [$\alpha$/Fe], T$_{\textup{eff}}$ and either L or $\Delta\nu$.
\\To explore potential systematics, we inferred masses and ages using two sets of constraints.
While T$_{\textup{eff}}$ and metallicity are included in both cases, in one set we considered as constraints  $\nu_{\textup{max}}$ and L, in the second  one $\nu_{\textup{max}}$ and $\Delta\nu$. As shown e.g. in \cite{Tailo2022}  datasets of relatively short duration  may be subject to systematics on the measurement of $\Delta\nu$; this is particularly relevant in the case of low-metallicity core-He-burning stars (see also \citealt{Matteuzzi2023}, Willett et al. in prep.).
In the set of ages inferred including $\Delta\nu$, we remove targets when their mass inferred  using ($\nu_{\textup{max}}$, $\Delta\nu$) differs from that determined by ($\nu_{\textup{max}}$, L) by more than 50\%. Given the direct impact on the high-$\alpha$ population, in the following we explore and check results and trends in both datasets.
\\Luminosities are computed using extinctions from the Bayestar19 dustmap (Green et al. 2014, 2019) implemented in the dustmaps python package (Green 2018) and bolometric corrections computed through the code by Casagrande $\&$ VandenBerg (2014, 2018a,b). We use the luminosity derived from $\overline{\omega}$+17$\mu$as, L$_{17}$. Distances are deduced from the \textit{Gaia} parallax, whereas in the ($\nu_{\textup{max}}$, $\Delta\nu$) sample they comes from PARAM.
\\When using PARAM, we considered a lower limit of 0.05 dex for the uncertainty on [Fe/H] and of 50 K for the uncertainty on T$_{\textup{eff}}$, due to the very low (internal) uncertainties quoted in APOGEE DR17 \citep[see also][]{Casali2023}. The grid of stellar models used here is the reference one adopted in the work of \cite{Miglio2021}, where a detailed explanation of the method can be found. 
\\We then compute the orbital parameters by using the fast orbit estimation method of \cite{Mackereth2018}, implemented in the GalPy package (Bovy 2015), where the Milky Way potential MWPotential2014 for the gravitational potential of the Milky Way is assumed  \citep{Bovy2015}. We assume that the radial position of the Sun is $R_{Gal,\bigodot}=8$ kpc and the circular velocity v$_{circ}=220$ km s$^{-1}$ \citep{Bovy2012}, the Sun's motion with respect to the local standard of rest $[U,V,W]_{\bigodot}=[-11.1,12.24,7.25]$ km s$^{-1}$ \citep{Schonrich2010}, and that the vertical offset of the Sun from the Galactic plane is $Z_{Gal,\bigodot}=20.8$ pc \citep{Bennett2019}.
\\The final reference sample considered here consists then of $\sim$ 6000 stars with available asteroseismic, spectroscopic and astrometric information. For further details on the observational sample, we address the reader to the catalogue paper by Willett et al. (in prep.).



\section{Are young $\alpha$-rich stars similar to old high-$\alpha$ populations?}


Here, we show the results for our sample of stars with available stellar ages from K2, chemical abundances from APOGEE and GALAH, and astrometric information from \textit{Gaia}. We start by defining the young $\alpha$-rich stars in our sample. Then, we discuss their chemical and kinematic properties. Our discussion is focused on APOGEE, and we will use the GALAH data to complement our analysis with respect to neutron-capture elements. 

\subsection{Definition}

\begin{table*}
	\centering
	\caption{Young $\alpha$-rich stars (YAR) in different bins of metallicity. In the first column, we indicate the considered range in [Fe/H].  In the second, third and fourth columns, we indicate the corresponding total number of stars, high-$\alpha$ stars and young $\alpha$-rich stars, respectively. In the fifth and sixth columns, we show the fraction of young $\alpha$-rich stars with respect to the total of stars and with respect to the high-$\alpha$ sequence, respectively. We report our results for ages computed with ($\nu_{\textup{max}}$, L) and, in parentheses, with ($\nu_{\textup{max}}$, $\Delta \nu$).}
	\label{Tab1}
 	\begin{tabular}{lcccccr} 
		\hline
		[Fe/H]		  & TOT & HA & YAR & f$_{TOT}$& f$_{HA}$\\
		\hline
		$>-$0.25 & 3335 (3021) & 344 (302) & 25 (24)  & 0.7$^{+0.1}_{-0.1}$\% (0.7$^{+0.2}_{-0.1}$\%)  & 7.3$^{+0.9}_{-1.2}$\%  (7.9$^{+2.0}_{-1.7}$\%)  \\
		$[-0.5,-0.25]$ & 2200 (1866)  & 1069 (875)  & 70 (85) & 3.2$^{+0.3}_{-0.2}$\% (4.6$^{+0.5}_{-0.3}$\%) & 6.5$^{+0.7}_{-0.4}$\% (9.7$^{+1.1}_{-0.7}$\%) \\
		$<-$0.5  & 1338 (998)  & 1115 (835)  & 73 (97)  & 5.5$^{+0.8}_{-0.4}$\% (9.7$^{+0.9}_{-0.9}$\%)  & 6.5$^{+1.0}_{-0.5}$\% (11.6$^{+1.0}_{-1.0}$\%) \\
		\hline
	\end{tabular}
 \end{table*}

First, we define the young $\alpha$-rich stars in our K2-APOGEE sample, similarly as done in \cite{Chiappini2015} with the CoRoGEE sample on the basis of the [Mg/Fe] vs. [Fe/H] plot (see Fig. 2). This is a well-known diagram to define young $\alpha$-rich stars \citep[see also more recently][]{Sun2020,Zhang2021,Cerqui2023}, but we now extend the asteroseismic study of \cite{Chiappini2015}
 by using our new K2 sample covering larger spatial baselines and with a much larger sample (10 times larger than the CoRoGEE one).
 \\In the upper left panel of Fig.~\ref{Fig3}, we show the region in the [Mg/Fe] vs. age plot that cannot be explained by means of chemical evolution models of the Galactic thin disc at different Galactocentric distances, as suggested in \cite{Chiappini2015}. The chemical evolution model is a multi-zone model for the Galactic thin disc at different Galactocentric distances, see \cite{Chiappini2009} and consistently the recent results of \cite{Grisoni2017,Grisoni2018}.
We then define the young $\alpha$-rich stars in our sample as being high-$\alpha$ stars ([Mg/Fe]>0.2 dex) and with young ages, well-inside the forbidden region defined by models. Namely, they should be inside the forbidden region and be at least 1 $\sigma_{age}$ away from the border of the selection region.
\\In the following sections, we will then compare the properties of our young $\alpha$-rich to other populations, namely the old high-$\alpha$ stars (i.e. the high-$\alpha$ stars not defined as young $\alpha$-rich) and the low-$\alpha$ ones. We divide our high- and low-$\alpha$ populations by considering a division in Mg, where the dichotomy is more evident \citep[see e.g. discussion in][]{Grisoni2017}. In particular, we consider [Mg/Fe]$>$0.2 and [Mg/Fe]$\le$0.2, respectively for the high- and low-$\alpha$ sequences. We perform this separation at [Mg/Fe] corresponding to 0.2 dex in order to obtain a "genuine" high-$\alpha$ sequence \citep{Miglio2021,Queiroz2023}, not including the transition/bridge region in our high-$\alpha$ definition \citep{Anders2018,Ciuca2021}.
\\We report the results in Table 1 and associate an uncertainty on the fraction of young $\alpha$-rich stars, by bootstrapping 1000 realizations of the data  by taking into account uncertainties in age and calculating the deviation of the distribution of the number of young $\alpha$-rich stars.  We find that the fraction of young $\alpha$-rich stars with respect to the high-$\alpha$ population is around 7-10$\%$ \citep[see also][]{Montalban2021,Miglio2021}. 
\\We note that the final fraction of young $\alpha$-rich stars can be affected by systematics. Comparing the ($\nu_{\textup{max}}$, L) and ($\nu_{\textup{max}}$, $\Delta\nu$) samples gives an estimate of potential systematics in the asteroseismic measurements. In particular, given our K2-APOGEE sample with all the required checks/flags, we tested also our second age dataset with ages from ($\nu_{\textup{max}}$, $\Delta\nu$); in this case, we get a slightly higher fraction of young $\alpha$-rich stars, but also in this case our main findings are not affected. We also tested other possible systematics in our definition of young $\alpha$-rich stars for our reference sample. First, we varied the threshold in the definition (without the 1 $\sigma_{age}$ constraint or with a stricter $2 \sigma_{age}$ constraint) and, in this case, the fraction of young $\alpha$-rich stars can vary from $\sim$5 to 15\%, but the spatial trends remain constant and not affected by the definition. We note also that, besides the threshold on the forbidden region, also the choice itself of the $\alpha$-element considered might affect the final results, but still within the range discussed in the aforementioned paragraphs (as previously discussed, here we choose Mg where the dichotomy between the two sequences is very evident). We stress that here a very large sample is available, but when doing such analyses with more biased samples \citep[see e.g.][]{Matsuno2018,Jofre2022} the selection of groups can be a limiting factor to be considered. We also note that the fraction of young $\alpha$-rich stars is higher in the red clump with respect to the red-giant branch, supporting the scenario in which most of these stars might have gone through interaction with a companion (see also \citealt{Miglio2021}).
\\In conclusion, even if the fraction of young $\alpha$-rich stars found in our sample can be slightly affected by the aforementioned systematics, the spatial trends across different parts of the Galaxy that we recover and discuss in detail in the next Sections are still preserved.

\subsection{Spatial properties}

\begin{figure*}
    \centering
    	\includegraphics[scale=0.35]{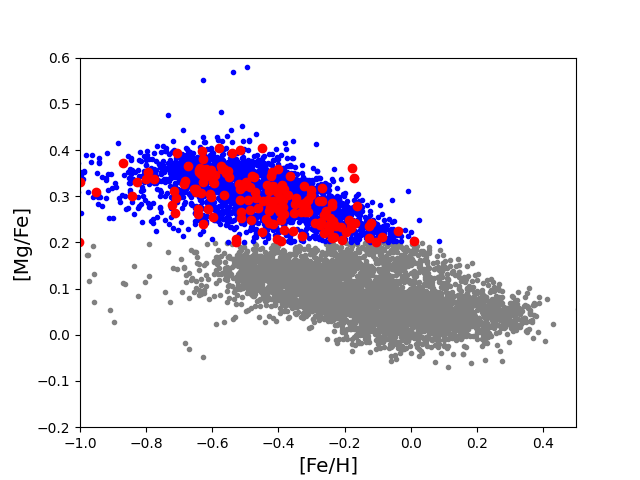}
 	\includegraphics[scale=0.35]{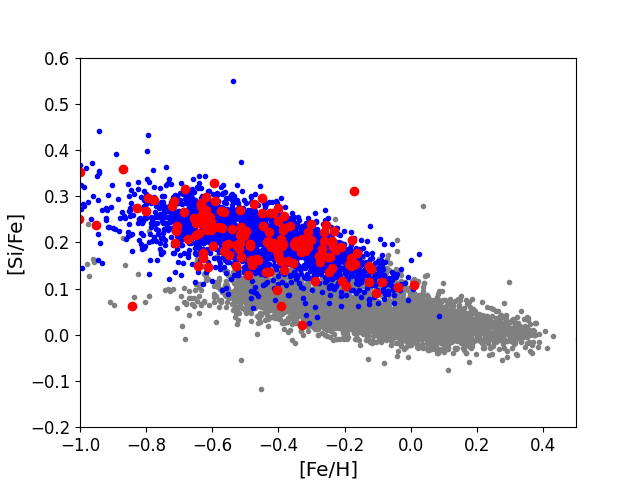}
   	\includegraphics[scale=0.35]{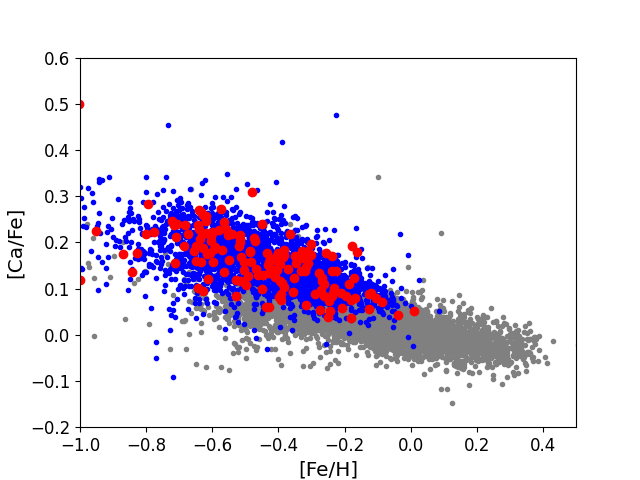}
    	\includegraphics[scale=0.35]{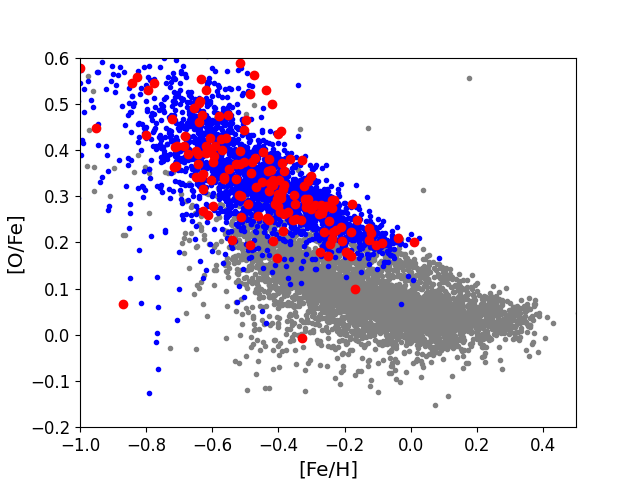}
     	\includegraphics[scale=0.35]{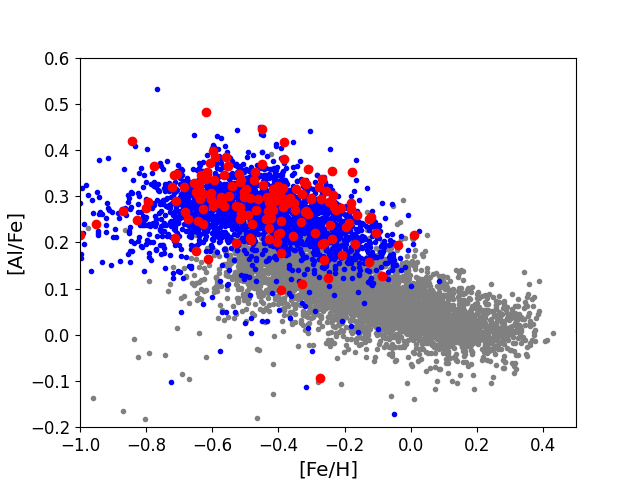}
   	\includegraphics[scale=0.35]{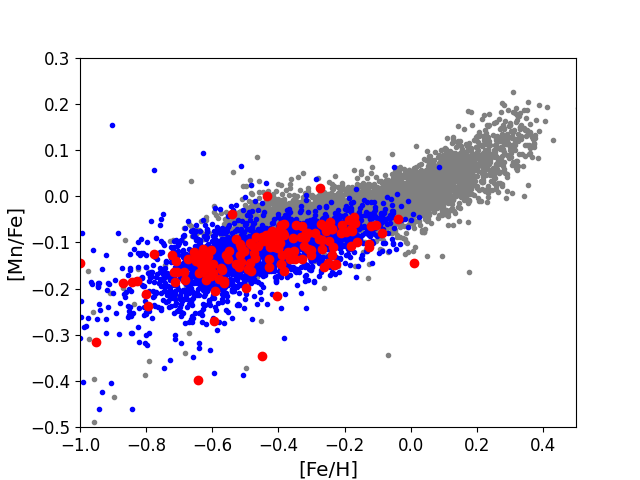}
                   	\includegraphics[scale=0.35]{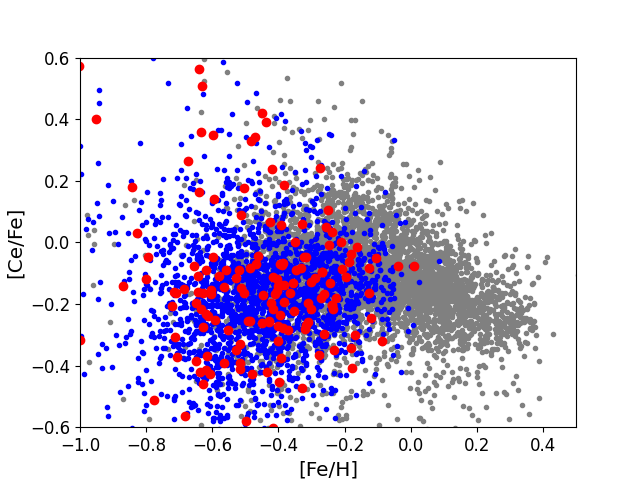}
       	\includegraphics[scale=0.35]{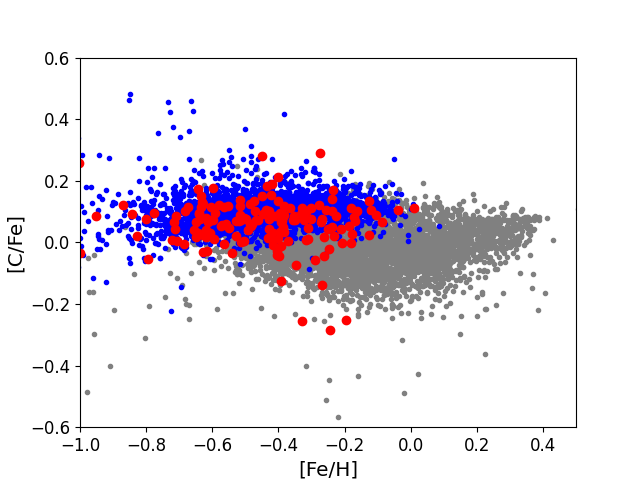}
           	\includegraphics[scale=0.35]{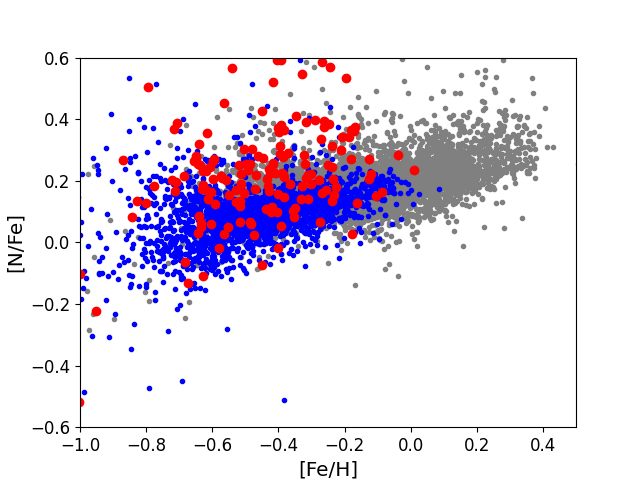}
    \caption{[X/Fe] vs. [Fe/H] plots for various chemical elements in our K2-APOGEE sample. Red dots represent the young $\alpha$-rich stars, blue dots the old high-$\alpha$ stars, and gray dots the low-$\alpha$ stars.}
    \label{Fig6}
\end{figure*}
 
Once our sample of young $\alpha$-rich stars has been defined, we study its spatial distribution in the Galaxy.
\\In the various panels of Fig.~\ref{Fig3}, we show the observational results  for our K2-APOGEE sample at different bins of guiding radius $R_g$ (the radius of a circular orbit of the same angular
momentum). In particular, we divide in the following bins of $R_g$ \citep[see also][]{Casali2023}: $R_g$ $<$ 5 kpc, 5-7 kpc, 7-8 kpc, 8-10 kpc, $R_g$ $>$ 10 kpc.  We make use of guiding radius rather than Galactocentric distance to mitigate blurring (see also \citealt{Willett2023}, and references therein).
\\From the various panels of Fig. 2, it is clear that similar fractions of young $\alpha$-rich stars are present in different parts of the Galaxy, from the innermost Galactic regions to the outer ones. As found in \cite{Chiappini2015} and \cite{Martig2015}, the fraction of young $\alpha$-rich stars with respect to the total number of stars is higher in the innermost part of the Galaxy, where the high-$\alpha$ sequence is dominating (see e.g. Table 1 in Chiappini et al. 2015). However, at variance with \cite{Chiappini2015} where young $\alpha$-rich stars seemed more abundant towards the inner Galactic disc regions
and therefore suggested the origin of these stars to be related to the complex chemical evolution that takes place near the co-rotation region of the Galactic bar, we find significant number of young $\alpha$-rich stars also in other bins of distance, where the high-$\alpha$ sequence extends. In particular, it is important to look at the fraction of young $\alpha$-rich stars with respect to the number of high-$\alpha$ stars, which remains almost constant across the Galaxy. With both our age samples, we find that the fraction of young $\alpha$-rich stars with respect to the total decreases going outwards in guiding radius, whereas the fraction with respect to the number of high-$\alpha$ stars remains almost constant within the errors in the different bins. 
\\Thus, to summarize, in our sample the young $\alpha$-rich stars appear in different Galactic locations, where the high-$\alpha$ sequence extends \citep{Hayden2015,Queiroz2020}. Therefore, this supports the idea that they can be part of the high-$\alpha$ population: they should have formed from the same gas as the high-$\alpha$ sequence, but they might be affected by binary evolution \citep{Izzard2018} 
\\Similarly, we investigate our results for the whole sample in three different bins in Galactic height (|Z| $<$0.5 , between 0.5-1 kpc, and $>$1 kpc).
The results concerning the number of young $\alpha$-rich found in these bins are summarized directly in Table 1. We can see that, in this case, the number of young $\alpha$-rich stars with respect to the total increases with height, as expected since the high-$\alpha$ sequence becomes dominant with respect to the low-$\alpha$ when going to greater height away from the Galactic plane \citep{Hayden2015,Queiroz2020}.
Still, the number of young $\alpha$-rich stars with respect to the high-$\alpha$ population remains almost constant, in agreement with what has been found as a function of guiding radius. Thus, the occurrence of young $\alpha$-rich stars does not depend on the location in the Galaxy, at variance with what was suggested by \cite{Chiappini2015} with CoRoGEE. This is found thanks to a much larger sample of stars in different parts of the Galaxy, which extends previous results found in the literature.

\subsection{Chemical properties}

\begin{figure*}
    \centering
    	\includegraphics[scale=0.35]{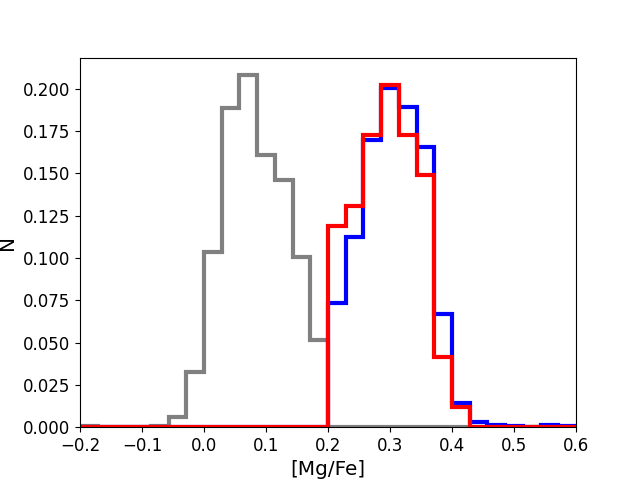}
 	\includegraphics[scale=0.35]{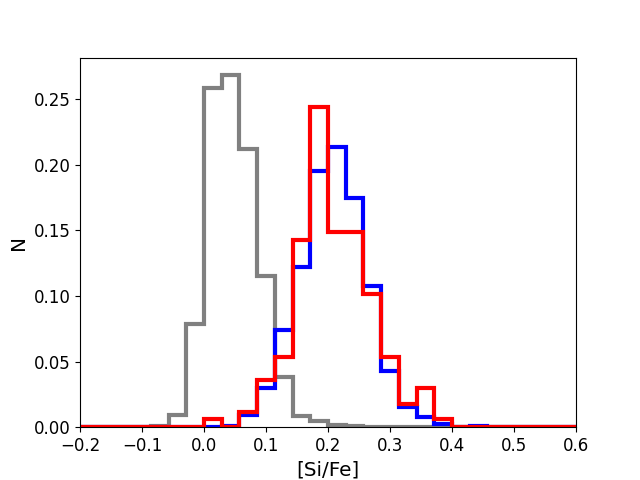}
   	\includegraphics[scale=0.35]{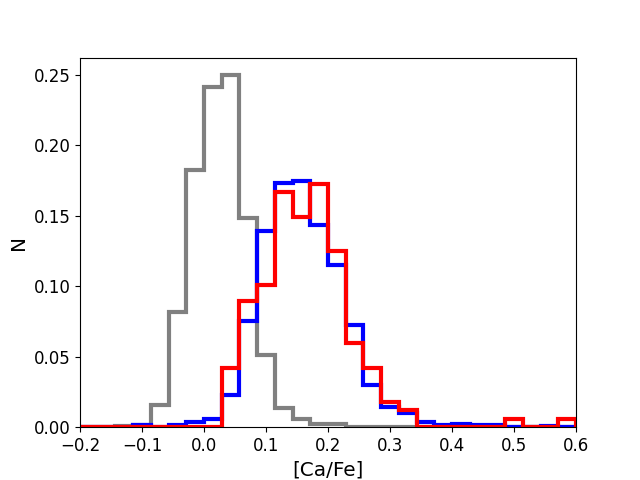}
    	\includegraphics[scale=0.35]{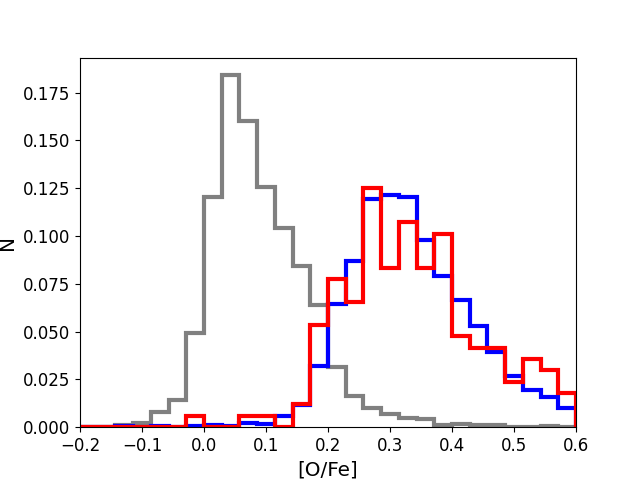}
     	\includegraphics[scale=0.35]{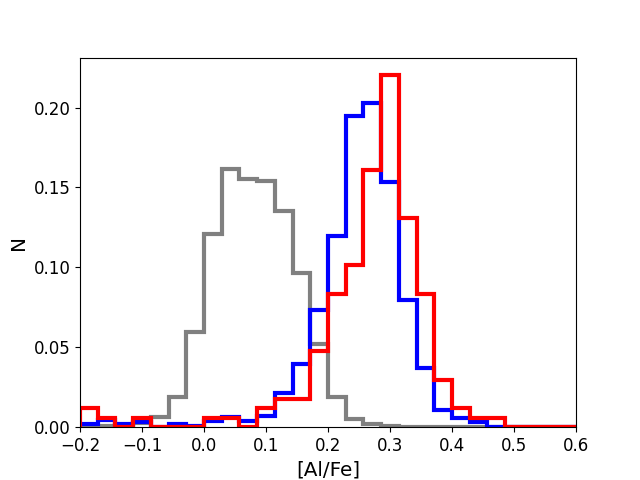}
   	\includegraphics[scale=0.35]{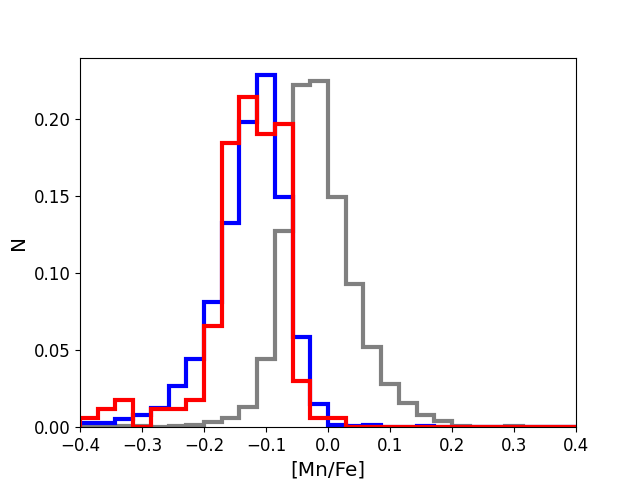}
    	\includegraphics[scale=0.35]{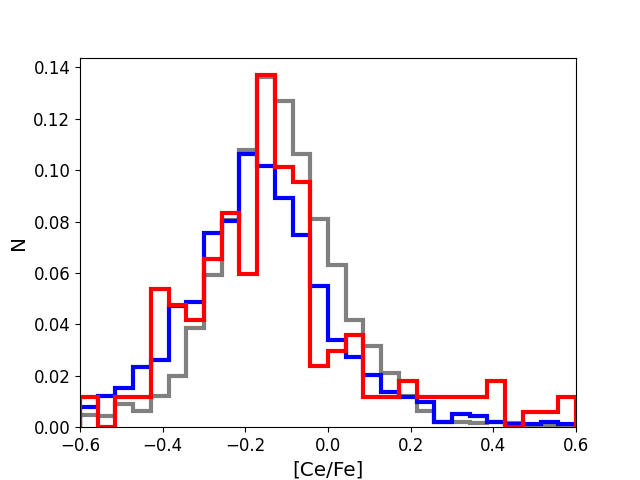}
       	\includegraphics[scale=0.35]{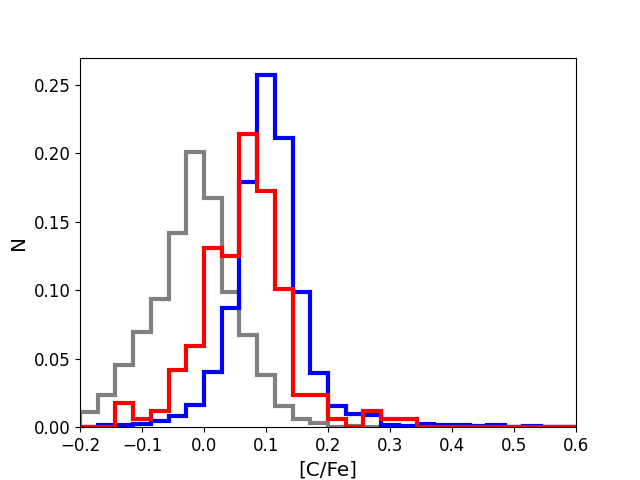}
           	\includegraphics[scale=0.35]{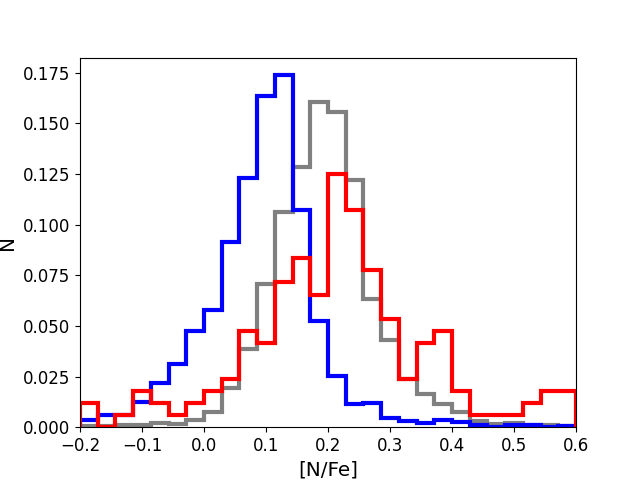}
    \caption{[X/Fe] vs. [Fe/H] plots for various chemical elements in our K2-APOGEE sample for the young $\alpha$-rich stars (in red), old high-$\alpha$ stars (in blue), and the low-$\alpha$ stars (in gray).}
    \label{Fig7}
\end{figure*}

\begin{figure*}
    \centering
    	\includegraphics[scale=0.35]{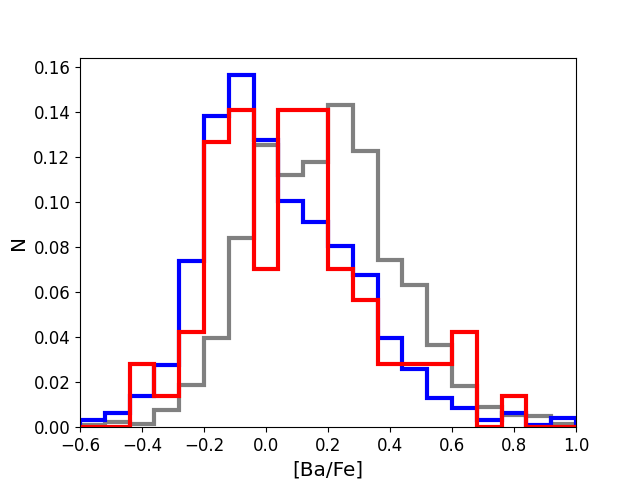}
 	\includegraphics[scale=0.35]{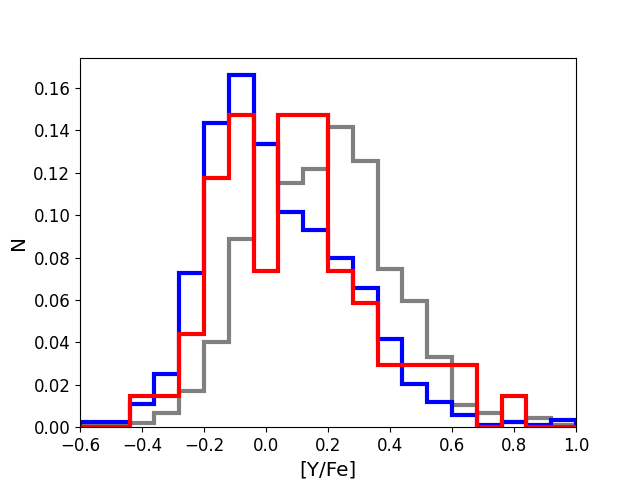}
   	\includegraphics[scale=0.35]{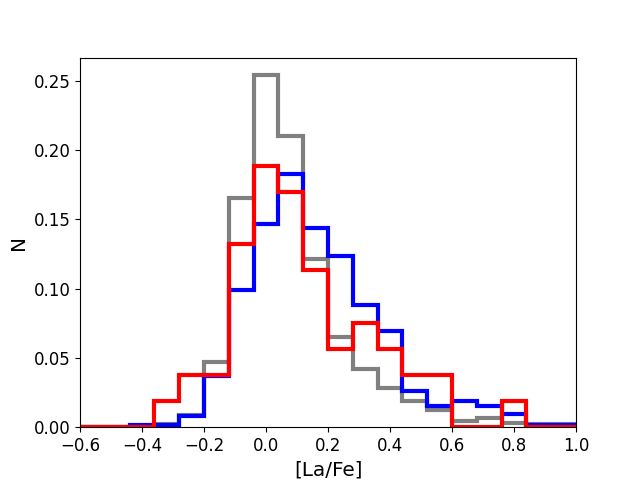}
    \caption{Same as Fig. 4, but for K2-GALAH.}
    \label{Fig7b}
\end{figure*}

\begin{figure*}
    \centering
    \includegraphics[scale=0.35]{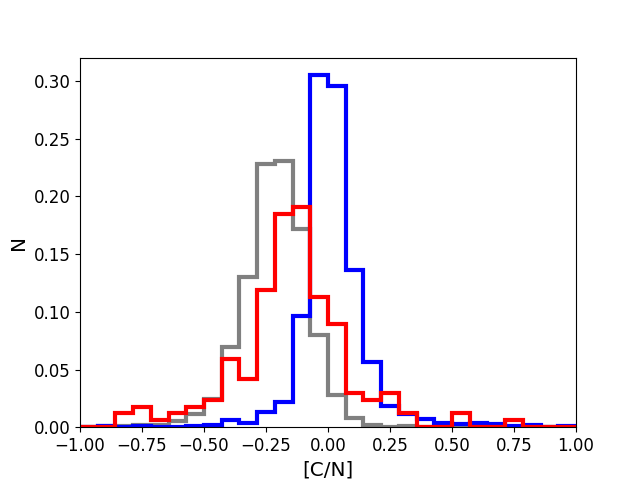}\
	\includegraphics[scale=0.35]{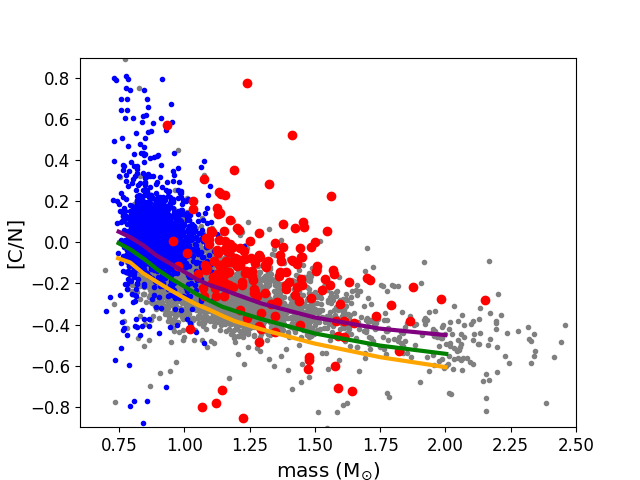}
 	\includegraphics[scale=0.35]{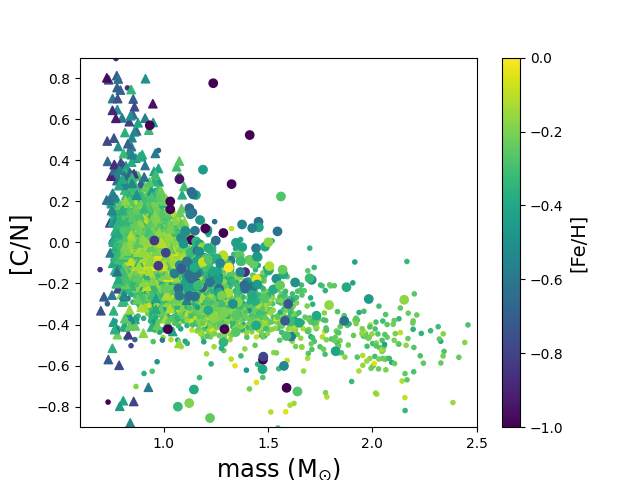}
    \caption{Left panel: histogram for [C/N] of the stars in our K2 sample (grey, blue and red histograms represent the low-$\alpha$, old high-$\alpha$ and young $\alpha$-rich samples, respectively). Middle panel: [C/N] versus mass for the stars of our sample, with colors as in the left panel. Solid lines represent predictions of stellar evolution models from \cite{Vincenzo2021} at different metallicities ([Fe/H]=-1 in yellow, -0.5 in green and +0.25 in purple, respectively). Right panel: [C/N] ratios for the entire sample color-coded by [Fe/H] (as indicated in the color scale). In this panel, we represent low-$\alpha$ (dots), old high-$\alpha$ (triangles), young-$\alpha$ rich stars (bigger dots). }
    \label{Fig8}
\end{figure*}

After showing the [Mg/Fe] vs. age plots used to define our sample of young $\alpha$-rich stars and investigate their occurrence rate in different parts of the Galaxy, we further investigate the chemical properties of our sample of young $\alpha$-rich stars by looking also at other chemical abundance patterns of different chemical elements considered reliable in APOGEE DR17 release \citep{APOGEEDR17}, and we will use the GALAH DR3 data \citep{Buder2021} to complement our analysis with respect to neutron-capture elements.

\subsubsection{Metallicity} 
We investigate the dependence on metallicity (i.e. [Fe/H]), by looking at different bins in [Fe/H] for our K2-APOGEE sample. The results of the various [Fe/H] bins are then directly reported in Table 2.
\\We find that young $\alpha$-rich stars are more dominant in metal-poor regimes, where the high-$\alpha$ sequence is dominating with respect to the total number of stars. With respect to the number of high-$\alpha$ stars, we still find a fraction of $\sim$7-10$\%$ for the young $\alpha$-rich stars, with a slightly increasing trend with decreasing metallicity though at the uncertainty level in the ($\nu_{max}$, $\Delta \nu$) sample, but this tentative trend is not there in the ($\nu_{max}$, L) sample.
We note that indeed there are evidences for an increased intrinsic fraction of close binaries in metal-poor regimes \citep[see][]{ElBadry2019,Moe2019} and an increased fraction of blue stragglers among old, metal-poor stars \citep{Fuhrmannetal2017,Casagrande2020}, with a more evident trend with decreasing metallicity with respect to what we find. Part of the difference could be due to the fact we likely removed (some) binary stars from our sample due to our selection criteria (e.g. \texttt{ruwe}).
\\Also, we find the fraction of our young $\alpha$-rich stars in the field to be similar to that of overmassive red giant stars found recently in open clusters, where they estimate an occurrence rate of around 10$\%$ and 5-10$\%$  in the old-open clusters NGC6819 and NGC6791 \citep{Brogaard2016,Handberg2017,Brogaard2021}, indicating that about 10$\%$ of the red giants in the cluster have experienced mass transfer or a merger.
\\These facts suggest that $\sim$10$\%$ of the low-$\alpha$ field stars could also have their ages underestimated by asteroseismology. This should be kept in mind when using asteroseismic ages to interpret results in Galactic archaeology.

\subsubsection{$\alpha$-elements}
Recently, also \cite{Jofre2022} and \cite{Cerqui2023} studied the nature of young $\alpha$-rich stars in the light of APOGEE data, but here we present new results based on stellar ages coming from asteroseismology and thus providing a complementary analysis with respect to the aforementioned works. In Fig.~\ref{Fig6}, we show our [X/Fe] vs. [Fe/H] plots for different chemical elements considered reliable in APOGEE DR17. 
\\We investigate the abundance patterns of the three populations introduced and defined at the beginning of Section 3, namely: young $\alpha$-rich stars (in red), old high-$\alpha$ (in blue), and low-$\alpha$ stars (as background in gray). After performing the separation among the three populations in the [Mg/Fe], we investigate also the various [X/Fe] vs. [Fe/H] diagrams for other chemical elements available and considered reliable in APOGEE DR17. We can see that, in general, the distribution of young $\alpha$-rich stars resembles the one of high-$\alpha$ stars rather than the low-$\alpha$ one, in agreement with previous studies. Thus, the young $\alpha$-rich stars share chemical properties more similar to the high-$\alpha$ population, in agreement with previous abundance analyses where they show indeed that the majority of the young $\alpha$-rich stars have chemical abundances similar to the high-$\alpha$ stars rather than the low-$\alpha$ stars at similar ages. Consistent results have been seen, for example, in \cite{Matsuno2018} using high-resolution spectroscopic follow up, and more recently in \cite{Jofre2022} and \cite{Cerqui2023} using also chemical abundances from APOGEE, but very different ages with respect to our asteroseismic estimates that can provide a complementary support to the idea that young $\alpha$-rich stars should be considered as stragglers of the thick disc, at variance with alternative scenarios present in the literature.
\\In the following, to be more quantitative, we then look at the corresponding histograms for the various [X/Fe] abundance ratios considered (similarly as done by \citealt{Zhang2023} with LAMOST data, but here for our APOGEE and GALAH data).
In Fig.~\ref{Fig7}, we show the corresponding histograms for the different chemical elements color-coded according to the different populations investigated in this work. In the case of Mg, the dichotomy between high and low-$\alpha$ stars is evident \citep[see e.g.][]{Grisoni2017}, and in fact we choose this chemical element to perform the separation between the high and low-$\alpha$ populations. For Mg, the young $\alpha$-rich stars are then part of the high-$\alpha$ population, by definition.
\\Then, we look at the other $\alpha$-elements, whether the separation is still evident and the young $\alpha$-rich stars follows the chemistry of the old high-$\alpha$ sequence. Also in the case of the other $\alpha$-elements considered here, we can see that in general the high and low-$\alpha$ populations can be well separated, and the young $\alpha$-rich stars share the same locus of the high-$\alpha$ population in the different diagrams.

\subsubsection{Aluminium} Aluminium is an odd-Z element. From Fig. 4, we can see that also in this case the histogram clearly shows a bimodal distribution, and that the young $\alpha$-rich stars follow well the high-$\alpha$ population.

\subsubsection{Manganese}
Mn is a Fe-peak element: in this case, the bimodality is not so clear, and high-$\alpha$ and low-$\alpha$ stars are not clearly separated. Still, the young $\alpha$-rich stars show a distribution in agreement with the high-$\alpha$ population. 

\begin{figure*}
\centering
 	\includegraphics[scale=0.35]{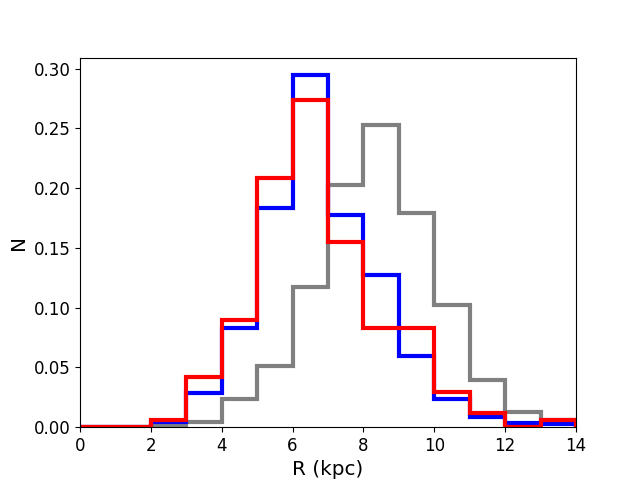} 
   	\includegraphics[scale=0.35]{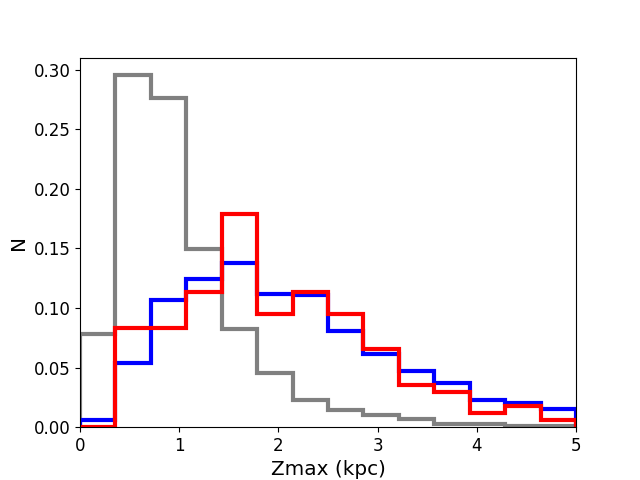} 
     	\includegraphics[scale=0.35]{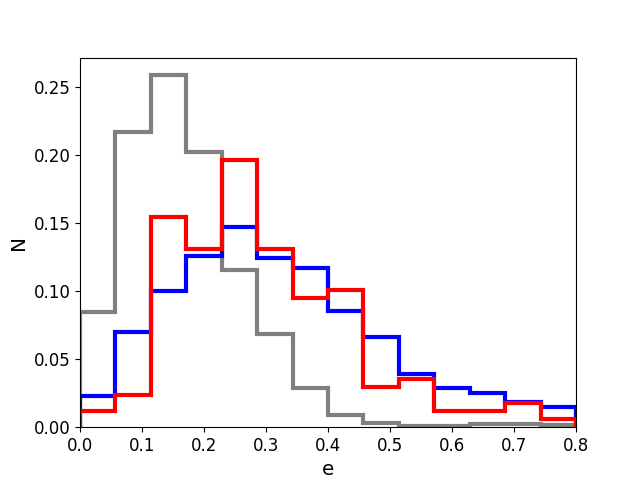} 
    \caption{Distribution function of guiding radii $R_g$ (left panel), $Z_{max}$ (middle panel) and eccentricity (right panel) for the different populations considered: young $\alpha$-rich (in red), old high-$\alpha$ (blue), low-$\alpha$ (grey).}
    \label{Fig9}
\end{figure*}

\subsubsection{Cerium}
Ce is the only s-process element available in APOGEE DR17 \citep[see][for a detailed study of the Ce abundances for our K2-APOGEE sample]{Casali2023}. In the case of Ce, the differences between the various populations are expected to be smaller, as predicted by chemical evolution models \citep[see e.g.][]{Grisoni2020a}. In the [Ce/Fe] vs. [Fe/H] plot (see Fig. 3), there is a large spread and it is difficult to draw firm conclusions whether young $\alpha$-rich stars are Ce-enhanced with respect the rest of the high-$\alpha$ sequence. In the histogram of Ce (see Fig. 4), the dichotomy between the various populations is not evident: the various populations are mixed and it is more difficult to disentangle among them. For example, \cite{Grisoni2020a} showed that the thick and thin disc populations are mixed in the abundance patterns of s-process elements such as Ce \citep[see also][]{Contursi2023} and it is more difficult to disentangle between them at variance with $\alpha$-elements, where the dichotomy between the two populations is evident. Since Ce is the only neutron-capture element available in APOGEE DR17 and it is difficult to draw firm conclusions, we decide to complement our analysis by performing a cross-match with GALAH DR3 in order to look at other neutron-capture elements.

\subsubsection{Other s-process elements with GALAH}


To better investigate neutron-capture elements, we took advantage of our K2-APOGEE sample cross-matched with GALAH DR3: in particular, we now consider the stars in our K2 sample that are in common between APOGEE and GALAH, where other s-process elements are available. For consistency, we mantain the same definition of young $\alpha$-rich stars as for the reference K2-APOGEE sample and complement it with information from GALAH. We have now a subsample of $\sim$2500 stars with also available abundances of neutron-capture elements from GALAH.
\\In particular, we show results for Y, Ba and La. Even if the scatter is higher with respect to APOGEE, this allows us to complement the analysis for neutron-capture elements not available in APOGEE.
\\From Fig. 5, we can see a mild s-process enhancement, but still it remains difficult to clearly assess whether there is a clear difference in s-process enhancement between the two populations (young $\alpha$-rich stars and old high-$\alpha$ stars). 
Therefore, further data on s-process elements are needed to establish whether the young $\alpha$-rich stars have peculiar properties regarding the abundance of s-process elements. In fact, looking at the abundance patterns of s-process elements can be important since they also can point towards some accretion of material. \cite{Zhang2021} took advantage of data from LAMOST survey and found that the young $\alpha$-rich stars were significantly Ba-enhanced compared to most of the high-$\alpha$ old stars: they explained the observed Ba-enhancement of the young $\alpha$-rich stars with the scenario that those stars are formed via binary evolution and, in particular, that they might have accreted Ba-rich materials from their AGB companions \citep[][]{Bidelman1951,Jorissen2019,Escorza2019,Zhang2023}. Such Ba enhancement of those young $\alpha$-rich stars was different from the findings of \cite{Yong2016} and \cite{Matsuno2018} that have shown that [Ba/Fe] ratios of the young $\alpha$-rich stars seem to be comparable to those of the old high-$\alpha$ stars. In the works of \cite{Yong2016} and \cite{Matsuno2018}, in fact it was found that Ba and other s-process elements were not systematically enhanced in young $\alpha$-rich stars and the reason might be related with the nature of the mass transfer, e.g. this happening before the primary reaches the AGB {\citep{Izzard2018,Jofre2022}}. Our results seem to confirm \cite{Matsuno2018} results with a larger sample, but further data might be needed to draw firm conclusions in this context.
\subsubsection{C and N}
Other chemical elements that can give very important hints about stellar evolution processes are C and N \citep[][for a detailed review]{Romano2022}.
\\From Fig. 3 and more evidently from the histograms in Fig. 4, we can see that for these elements there are clear differences between the old high-$\alpha$ and the young $\alpha$-rich stars, especially for N. We remind the reader that the surface abundances of C and N are affected by the first dredge-up, with higher mass stars dredging up material with increased He and N and decreased C and Li \citep{Salaris2015,Salaris2018}. The changes in C and N are thus a result of the burning in the interior, which is dominated by the CNO cycle. The offset in C and N for some of the young $\alpha$-rich stars most probably arises from the larger masses of the young $\alpha$-rich population \citep[see also][]{Hekker2019}. From our data, we can also see that young $\alpha$-rich stars seem to present slightly lower C abundances and more evidently higher N abundances with respect to the old high-$\alpha$ stars. We note the presence of many N-rich stars in our sample \citep[see also][]{Schiavon2017,Trincado2022}{}{}. 
\\Differences in C and N are then reflected in the [C/N] ratio itself.
Further insights on the nature of the young $\alpha$-rich stars stars can thus be gained from the [C/N] ratio \citep[see e.g.][]{Hekker2019,Jofre2016,Jofre2022,Izzard2018,Sun2020,Miglio2021,Cerqui2023}.
In Fig.~\ref{Fig8}, we focus on the [C/N] ratios of the stars in our sample. As can be seen from the histogram (left panel), the young-$\alpha$-rich stars (red line) lie closer to the low-$\alpha$ (grey line), differing substantially from the distribution of the old high-$\alpha$ stars (blue line). Since part of this difference is expected due to the higher masses of the young-$\alpha$-rich stars compared to old high-$\alpha$ stars, at least one more dimension is needed to identify potential signs of binary evolution. Therefore, in the other two panels in the figure, we show the [C/N] ratio as a function of mass. In the
middle panel, the three samples are shown (low-$\alpha$, high-$\alpha$ stars, young-$\alpha$-rich, colored as in
the left panel). We also overplot three curves representing
stellar model predictions from \cite{Vincenzo2021} of the dependency of the [C/N] ratio with stellar mass. The purpose is to illustrate what has been said above, i.e., that lower ratios are expected for larger masses. However, the curves are not meant to fit the data, as here there is a complex mix of stellar populations (see e.g. discussion in \citealt{Anders2018}), and thus it is out of the scope of the present work to identify debris from past accretion, debris from globular clusters, or stars migrating from their original birth radii.
Focusing on the stellar evolution aspects only, we see a
larger spread in the [C/N] ratio at a given mass for the young-$\alpha$-rich stars (red) with respect to what is seen in the low-$\alpha$ sample (grey). The young-$\alpha$-rich stars with [C/N] values above the general trend can arise if the current star is the result of a merger or mass-transfer between two lower-mass giants, which have preserved their original [C/N] values. A more detailed explanation is given in \citet{Jofre2016} where their Fig. 5 shows a simulation of [C/N] vs. age for a stellar population including binary evolution. The simulation predicts that many of the interacting binaries result in a sequence of stars where [C/N] does not depend on mass and remains high independently of the mass. Though we have relatively few such stars, we do see a number of stars that scatter close to line [C/N]=0 line above the general single-star trend, which are consistent with this scenario. Importantly, these conclusions remain valid even if one considers shifting the single-star theoretical predictions in the middle panel of Fig.~\ref{Fig8} upwards to follow the upper envelope of the observed low-$\alpha$ star sequence. Some of the $\alpha$-rich stars show very high [C/N] ratios. These large ratios are also seen in some of the
high-$\alpha$ sample (blue). In the right panel, where we color our samples with metallicity, it can be seen that most of the stars with very large [C/N] ratios tend to be more metal-poor (C-enhanced stars). The high [C/N] values for the young-$\alpha$-rich stars can thus also be a hint of binary mergers/mass accretion, in this case for stars that started out with higher abundances of C. 
\\Further observational evidence would be important to clearly assess the nature of these stars, but to the best of our knowledge there is currently no other explanations for the [C/N] values higher than the single-star trend
while they are consistent with simulations in the case of binary mergers/mass-transfer.
\\

\subsection{Kinematic properties} 
In this section, we look at the kinematics of these young $\alpha$-rich stars and address the question of whether the old high-$\alpha$ and the young $\alpha$-rich stars differ in kinematics.
\\In Fig.~\ref{Fig9}, we show the distribution of guiding radius $R_g$, $Z_{max}$ and eccentricity for the three different populations (young $\alpha$-rich, old high-$\alpha$, and low-$\alpha$). From this figure, we can see that the young $\alpha$-rich stars share similar properties with the old high-$\alpha$ population: they are vertically hotter than the low-$\alpha$ population and they display properties in agreement with the distribution of the rest of the old high-$\alpha$ sequence both in $R_g$, $Z_{max}$ and eccentricity. This seems to be inconsistent with previous interpretations of the spatial distribution of these young $\alpha$-rich stars as coming from the inner part of
the Galaxy \citep{Chiappini2015}, suggesting that they formed in the bar region and migrated outwards. The latter scenario could be the result of an incomplete sampling in previous studies. Conversely, as found by many other studies \citep{Martig2015,SilvaAguirre2018}, we can see that the kinematics of our young $\alpha$-rich stars is more similar to old high-$\alpha$ stars rather than the low-$\alpha$ stars with similarly young ages. These findings support the scenario in which the majority of these young $\alpha$-rich stars share the same properties of the genuine high-$\alpha$ population, and thus the idea that they are part of the high-$\alpha$ sequence but probably they are products of mergers/mass accretion from a companion star \citep{Martig2015,Yong2016,Jofre2016,Jofre2022,Zhang2021,Miglio2021}. Consistent trends were found also by \cite{Sun2020} and discussed also in \cite{Jofre2022,Cerqui2023}. They all conclude that young $\alpha$-rich stars should be considered as stragglers of the thick disc, at variance with alternative scenarios proposed in the literature. Our results can thus complement and support these conclusions with a new sample accounting for precise ages coming from asteroseismology.
\\To summarize, with our new K2 sample we are able to study young $\alpha$-rich stars in different parts across the Galaxy and investigate in detail their spatial distribution, chemical properties and kinematic properties. We find that these stars are present in every bin of Galactocentric distance and Galactic heights, where the high-$\alpha$ sequence is present. They present both chemical and kinematic properties more similar to the high-$\alpha$ stars rather than the low-$\alpha$ stars at the same age. Therefore, these young $\alpha$-rich stars are like high-$\alpha$ but might be the product of binary evolution with merger/accretion, in agreement with previous studies \citep{Yong2016,Jofre2016,Jofre2022,Zhang2021} rather than related to a peculiar chemical evolution scenario near the co-rotation region \citep{Chiappini2015}.


\section{Summary and conclusions}

In this paper, we have investigated the nature of young $\alpha$-rich stars in an unprecedented sample of $\sim$ 6000 red giants observed with K2, with spectroscopic information from APOGEE DR17 and GALAH DR3, then cross-matched with \textit{Gaia}. Our new K2 sample spans a wider range of locations in the Galaxy and thus allows to perform a novel more comprehensive analysis of young $\alpha$-rich stars in the Galaxy with respect to previous asteroseismic studies in the literature, such as the one of \cite{Chiappini2015} with CoRoGEE. Moreover, with its precise asteroseismic ages, it allows to complement other recent studies of young $\alpha$-rich stars \citep[see e.g.][]{Jofre2022,Cerqui2023}, whose nature is still very much debated in Galactic archaeology.
\\Our main conclusions are as follows.
\begin{itemize}
    \item By applying the definition of young $\alpha$-rich stars in the [Mg/Fe] vs. age plot, we define our sample of young $\alpha$-rich stars. We find that the fraction of young $\alpha$-rich stars with respect to the high-$\alpha$ population is around 7-10$\%$ \citep[see also][]{Montalban2021,Miglio2021}, and we discuss possible systematics affecting this fraction. 
    \item The young $\alpha$-rich stars present in our sample are found in each bin of $R_g$ (from less than 4.5 kpc to above 10.5 kpc). The percentage of young $\alpha$-rich stars with respect to the total number of stars decreases with guiding radius since the high-$\alpha$ sequence becomes less dominant; however, the percentage of young $\alpha$-rich stars with respect to the number of high-$\alpha$ stars remains constant.
    \item Similarly, we investigate the presence of young $\alpha$-rich stars in different bins of |Z| and find that they become more dominant at higher |Z| since the high-$\alpha$ sequence becomes more dominant; still, the fraction of young $\alpha$-rich stars with respect to the high-$\alpha$ population stays constant ($\sim$ 7-10$\%$ ). Thus, independently from the definition, we highlight the fact that the fraction of young $\alpha$-rich stars with respect to the number of high-$\alpha$ stars remains almost constant across different parts of the Galaxy.
    \item If the young $\alpha$-rich stars are interpreted as having gained mass through binary evolution, our findings are also in agreement with studies of open clusters, where an occurrence rate of overmassive red giant stars of about 10$\%$ has been also found \citep{Handberg2017,Brogaard2018,Brogaard2021}.
    \item Concerning the chemical properties, we have considered APOGEE DR17 chemical abundances and found that young $\alpha$-rich stars share similar chemical abundances to those of the old high-$\alpha$ population, except for elements such as C and N.
    \item We have also used GALAH DR3 chemical abundances  to complement our analysis with respect to neutron-capture elements not available in APOGEE. In particular, we considered Y, Ba and La, and found a mild s-process enhancement. Still, it remained not possible to assess whether there are clear differences in s-process elements between young $\alpha$-rich stars and the old high-$\alpha$ ones, and new data of higher precision would be valuable.
    \item Also regarding the kinematic properties of young $\alpha$-rich stars in our sample, they look similar to those of the high-$\alpha$ sequence rather than the low-$\alpha$ one.
\end{itemize}
With our new K2 sample that spans a wider range of Galactocentric distances than before, we conclude that young-$\alpha$ rich stars are present
in different parts of the Galaxy and they share the same properties as the normal $\alpha$-rich population, except for [C/N]. Binary evolution with accretion and/or mergers are naturally consistent with these findings.

\begin{acknowledgements}
We thank the referee for useful comments and suggestions that improved our paper.
\\VG, CC, AM, KB, GC, EW, JM, AS, JST, MT, MM acknowledge support from the ERC Consolidator Grant funding scheme (project ASTEROCHRONOMETRY, https://www.asterochronometry.eu, G.A. n. 772293). VG and CC acknowledge support from the EU COST Action CA16117 (ChETEC). CC acknowleges  Fundaci\'on Jes\'us Serra for its great support during her visit to IAC, Spain, during which part of this work was written. Funding for the Stellar Astrophysics Centre is provided by The Danish National Research Foundation (Grant agreement No.~DNRF106). YE acknowledges support from the School of Physics and Astronomy, University of Birmingham. VG also acknowledges financial support from INAF under the program “Giovani Astrofisiche ed Astrofisici di Eccellenza - IAF: Astrophysics Fellowships in Italy" (Project: GalacticA, "Galactic Archaeology: reconstructing the history of the Galaxy").

\end{acknowledgements}

\bibliographystyle{aa}
\bibliography{aa_biblio}

\begin{appendix}

\section{IDs for young-$\alpha$ rich stars}

Here, we show the example of list of IDs for the best-candidate young-$\alpha$ rich stars, that we are planning to make available online with the present work. All the asteroseismic, spectroscopic and astrometric information will be then released together with the catalogue paper (Willett et al. in prep.).

\begin{table*}
\caption{Best-candidate young $\alpha$-rich stars (YAR) in our sample. We report the K2 ID and the corresponding K2 campaign.}
\label{Tab1}
  \begin{tabular}{l} 
\hline
IDs\\
\hline
KTWO212300275-C06\\
KTWO213437162-C07\\
KTWO212626515-C06\\
KTWO201885410-C01\\
KTWO246472224-C12\\
...\\
\end{tabular}
 \end{table*}

\end{appendix}

\end{document}